# Electrically inert *h*-BN/bilayer graphene interface in all-2D-heterostructure FETs

*Teerayut Uwanno,*[†,‡] *Takashi Taniguchi,*[§] *Kenji Watanabe,*[§] *and Kosuke Nagashio*[*,†,∥]

[†]Department of Materials Engineering, The University of Tokyo, Tokyo 113-8656, Japan
[‡]College of Nanotechnology, King Mongkut's Institute of Technology Ladkrabang, Bangkok 10520, Thailand
[§]National Institute of Materials Science, Ibaraki 305-0044, Japan
[∥]PRESTO, Japan Science and Technology Agency (JST), Tokyo 113-8656, Japan
E-mail: nagashio@material.t.u-tokyo.ac.jp

**Abstract**
Bilayer graphene field-effect transistors (BLG-FETs), unlike conventional semiconductors, are greatly sensitive to potential fluctuations due to the charged impurities in high-*k* gate stacks since the potential difference between two layers induced by the external perpendicular electrical filed is the physical origin behind the band gap opening. The assembly of BLG with layered *h*-BN insulators into van der Waals heterostructure has been widely recognized to achieve the superior electrical transport properties. However, the carrier response properties at the *h*-BN/BLG heterointerface, which control the device performance, have not yet been revealed due to the inevitably large parasitic capacitance. In this study, the significant reduction of potential fluctuations to ~1 meV is achieved in all-2-dimensional heterostructure BLG-FET on a quartz substrate, which results in the suppression of the off-current to the measurement limit at a small band gap of ~90 meV at 20 K. By capacitance measurement, we demonstrate that the electron trap/detrap response at such heterointerface is suppressed to undetectable level in the measurement frequency range. The electrically inert van der Waals heterointerface paves the way for the realization of future BLG electronics applications.
**KEYWORDS:** bilayer graphene, band gap, potential fluctuation, FET, heterostructure

## INTRODUCTION

The main factors that limit carrier modulation in metal-oxide-semiconductor field effect transistors (MOSFETs) are interface traps and potential fluctuations due to fixed charges in oxide insulators.[1] In conventional semiconductors such as Si, SiC and other compound semiconductors, the intrinsic band gap of the channel materials is sufficiently large compared to the potential fluctuation of 30~120 meV;[2-5] therefore, the main strategy to improve the device performance has been the reduction of the interface trap density ($D_{it}$).

On the other hand, bilayer graphene (BLG), a promising high-mobility channel, possesses an electrostatically tunable but relatively small band gap ($E_g$) up to ~0.3 eV attained by applying a displacement field ($\bar{D}$).[6, 7] The displacement field induces different carrier densities in the top and bottom layers of BLG, resulting in potential difference between the two layers. This inversion symmetry breaking is the physical origin behind the band gap opening.[8] Therefore, the conventional high-*k* oxide/BLG gate stack on a SiO$_2$/Si substrate, which introduces charge inhomogeneity with potential fluctuations of 20~30 meV,[9, 10] has suffered from high off-current ($I_{off}$) due to the spatial variation of the band gap in the channel, which leads to a low on/off current ratio ($I_{on}/I_{off}$).[11-20] Moreover, contrary to the transport measurement, the capacitance-voltage (*C-V*) measurement is often used to measure the mobile carrier response to a small-signal alternating current voltage at a certain frequency to elucidate the interface properties. Since the charge inhomogeneity in the oxide gate stack introduces the disorder in the gap, the gap state density has been estimated to be ~$10^{13}$ cm$^{-2}$eV$^{-1}$ by conductance method in high quality high-*k* Y$_2$O$_3$ top gate insulators,[19] which has a value that is two orders higher than that in SiO$_2$/Si system. Therefore, the predominant issue for BLG is to reduce the charged impurities, unlike conventional semiconductor systems.

Since *h*-BN was first recognized as a substrate for achieving high carrier mobility in graphene due to reduced charged impurities and atomically flat surface,[21] there have been many reports on the improved transport properties for BLG/*h*-BN heterostructures on SiO$_2$/Si wafers.[22-34] Because of the high mobility and reduced disorder obtained, topological issues have recently been intensively investigated through the electrical transport properties,[35-37] and topological edge currents were proposed as an alternative origin, rather than gap states to explain why $I_{off}$ often saturates around low $\bar{D}$ (~0.2-0.3 V/nm) even in the *h*-BN-encapsulated BLG. Indeed, the edge current was reported in *h*-BN-encapsulated BLG on a SiO$_2$/Si wafer with the Hall bar geometry, while no conduction path was found in the edgeless Corbino geometry.[36] However, it is still under debate because the zigzag edge structure expected for the edge current is generally difficult to form and the contribution of remnant charged impurities is still suggested. Here, in order to screen the charged impurities on the SiO$_2$/Si substrate, the h-BN substrate itself is not enough because the global back gate propagates the charge inhomogeneity in oxides through h-BN to BLG, which may lead to non-ideal suppression of $I_{off}$. Therefore, graphite back gate electrodes have often been used recently.[36, 38-41]

Despite the common understanding of the superior electrical transport properties of h-BN-encapsulated BLG, the carrier response properties at the h-BN/BLG heterointerface, which control the device performance, have not yet been revealed due to the inevitably large parasitic capacitance ($C_{Para}$) from the high-doped Si substrate as the global back gate. In this study, the charged impurities and $C_{Para}$ are reduced to the limit by using graphite as both back and top gate electrodes for h-BN-encapsulated BLG on a high insulating quartz substrate. Using this all-2-dimensional (2D) heterostructure BLG-FET on a quartz substrate, we demonstrate the significant reduction of potential fluctuations to ~1 meV, which results in the suppression of the off current to the measurement limit at a small band gap of ~90 meV at 20 K. The frequency-dependent C-V measurements revealed no carrier response in the measurement range up to 2 MHz, suggesting that the h-BN/BLG heterointerface in all-2D structures is electrically quite inert.

## RESULTS
### All-2D heterostructure assembly.

All-2D heterostructures were fabricated by a method similar to previously reported pick up method using polymers of polydimethylsiloxane (PDMS) and polymethylmethacrylate (PMMA).[42-44] Detailed procedure can be seen in **Note 1**. Using a micromanipulator system,[44] van der Waals heterostructure composed of h-BN top gate, BLG, h-BN back gate and graphite back gate electrode were fabricated onto a quartz substrate. To obtain a clean BLG/h-BN interface, bubbles containing air and hydrocarbons must be removed. However, once BLG is encapsulated by h-BN crystals, interfacial bubbles did not aggregate with the subsequent annealing, unlike BLG on h-BN.[44] The selection of BLG with a width narrower than 2 μm resulted in fewer bubbles being trapped in BLG channel area during encapsulation, as shown in **Fig. S2**. Moreover, it was observed that bubbles formed near the edge of the BLG flake encapsulated by h-BN crystals often migrate out of the BLG flake after annealing, making it possible to obtain a larger clean interface area in narrow BLG for further device fabrication.

Because BLG is completely encapsulated in h-BN, electrical contacts cannot be directly formed on BLG. In previous studies, the contact region was entirely etched by CHF$_3$/O$_2$ plasma with 60 W power in a tilted profile to form edge contact with metal electrodes.[42] In this study, it was found that h-BN can be selectively etched by CF$_4$ plasma when the power is reduced to 18 W. The high selectivity of 1:100 for BLG and h-BN is due to the chemical reaction being more dominant than physical etching in low power plasma. The detailed electrode formation process is shown in **Fig. S3**. It should be emphasized that the yield for ohmic contacts is greater than 80 % because of the surface contact.

Metal top gate electrodes have been

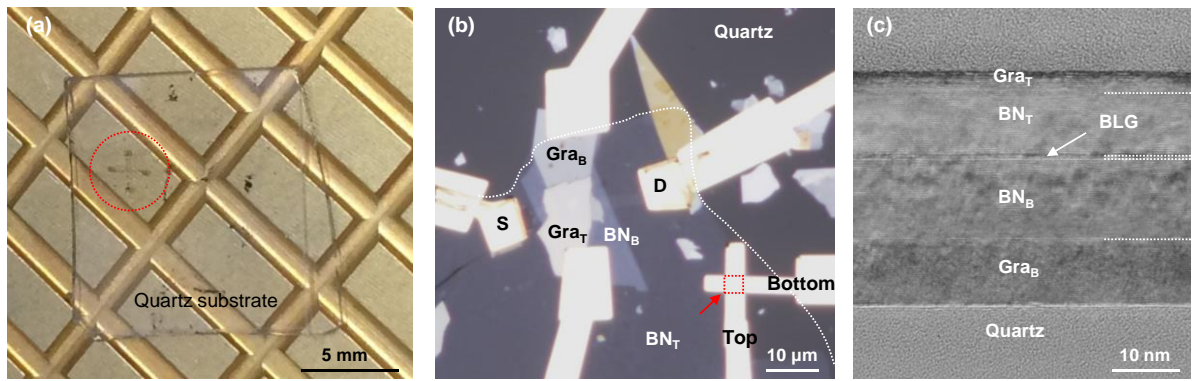

**Figure 1** (a) Optical image of an all-2D heterostructure BLG-FET on a quartz substrate. (b) Enlarge image of the circled area in (A). The symbols used in the figure are as follows. The graphite top gate electrode (Gra$_T$), h-BN top gate (BN$_T$), h-BN back gate (BN$_B$), and graphite back gate electrode (Gra$_B$). The large BN$_T$ was intentionally used, whose edge is indicated by a white dotted line. A metal/BN$_T$/metal parallel plate capacitor was fabricated at the position shown by a red arrow to measure the geometric capacitance of BN$_T$. Due to the present source/drain electrode fabrication process, graphite top and back gate electrodes does not overlap the source/drain electrodes. (c) Cross sectional TEM image of a typical all-2D heterostructure BLG-FET. The thicknesses of Gra$_T$, BN$_T$, BN$_B$, and Gra$_B$ are 5 nm, 8.7 nm, 10 nm, and 11.6 nm, respectively.

conventionally used in BLG devices so far. However, electron back scatter diffraction pattern (EBSD) analysis revealed that the Ni/Au top gate metal electrode was tiny polycrystalline, leading to an increase in the potential fluctuation in BLG due to the variation of the work function for different crystal orientations, as shown in **Fig. S4**. Therefore, to minimize the potential fluctuations in BLG, graphite was also used as the top gate electrode for $h$-BN-encapsulated BLG,[45] creating all-2D heterostructure, as shown in **Fig. 1**. The cross sectional transmission electron microscope (TEM) image (**Fig. 1(c)**) indicates that the interface between $h$-BN and BLG is clean and atomically sharp with a surface roughness (RMS) of <0.1 nm, which is confirmed by an atomic force microscope (AFM).

**Drastic improvement of $I_{on}/I_{off}$.**

To evaluate $I_{on}/I_{off}$ for all-2D heterostructure BLG-FET, two terminal conductivity was measured as a function of $V_{TG}$ for each $V_{BG}$ at 20 K, as shown in **Fig. 2(a)**. A quite sharp ambipolar behavior was observed. As shown in the enlarged figure, $I_{off}$ was suppressed to the measurement limit at $\bar{D}$ < -1.34 V/nm, where the resistance was larger than ~5 GΩ. This is quite similar with the recent report[46] and exceeds previously reported values of ~10-100 kΩ.[14-16,19] $I_{off}$ in all-2D heterostructure BLG was reduced compared to $h$-BN-encapsulated BLG on SiO$_2$/Si as well as high-$k$ Y$_2$O$_3$/BLG on SiO$_2$/Si at the same $|\bar{D}|$, as shown in **Fig. 2(b)**, indicating drastic reduction of the charged impurities and hence the spatial uniformity of the band gap of BLG. As a result, $I_{on}/I_{off}$ is greatly increased compared to other devices at the same $|\bar{D}|$, and the maximum $I_{on}/I_{off}$ at 20 K was 4.6×10$^5$ at $\bar{D}$ = -1.48 V/nm, which is the best data thus far for $h$-BN-encapsulated BLG-FETs. Another important advantage of using graphite as back gate electrode is that the maximum $\bar{D}$ attained is increased due to the utilization of thin $h$-BN compared to that in the $n^+$-Si global back gate, because the total 2D crystal thickness of ~20 nm required to eliminate the SiO$_2$

**FIG. 2**

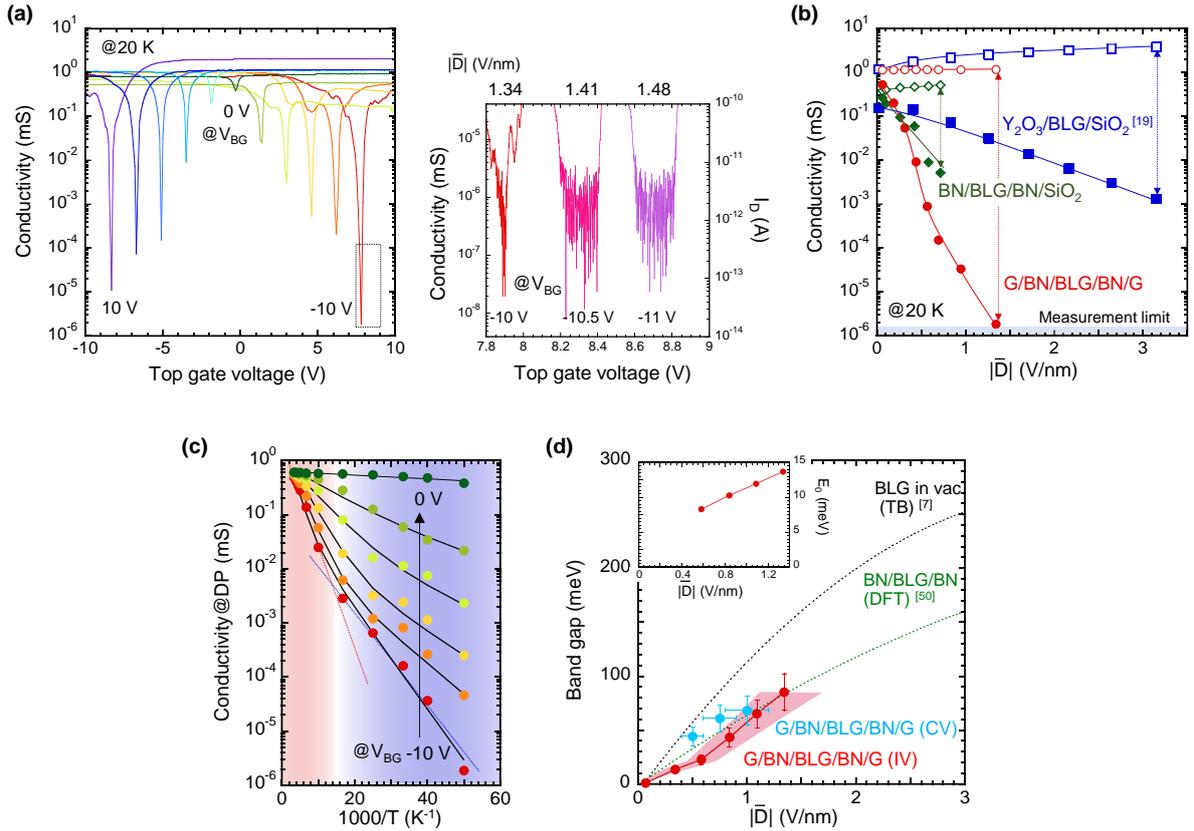

**Figure 2** (a) Left: two-terminal conductivity as a function of top gate voltage for each back gate voltage measured at 20 K. Right: drain current as a function of top gate voltage for back gate voltages of -10 to -11 V at $V_D$ = 0.01 V. (b) Conductivity for $I_{on}$ and $I_{off}$ as functions of $|\bar{D}|$ measured at 20 K. The open and closed symbols indicate $I_{on}$ and $I_{off}$ for different device structures, respectively. (c) Temperature dependence of conductivity at the Dirac point for different $V_{BG}$. (d) Band gap and $E_0$ extracted from (c) and $C$-$V$ measurements as a function of $|\bar{D}|$. The colored region for the band gap estimated from the $I$-$V$ measurement represents the range of the result from using the dielectric constant of 3.5-4.5 for $h$-BN to calculate $|\bar{D}|$.

surface roughness can be mainly covered by the graphite back gate. The field effect mobility was estimated to be ~20000 cm$^2$/Vs at 20 K. The subthreshold swing was 40 mV/dec at 20 K, suggesting the improvement in interface properties. The transfer characteristics with forward and backward top gate voltage sweeps at 20 K are shown in **Fig. S5**. The hysteresis is negligible and the device shows exactly the same performance and charge neutrality point even at 1 year after the fabrication.[47] Minor peaks observed near the Dirac point for each back gate voltage at lower temperature might be formed by moire potential from $h$-BN.[24]

The temperature dependence of the conductivity at the Dirac point (σ@DP) shows two activation processes in measured temperature range, as shown in **Fig. 2(c)**. They are known as thermal activation (TA) at high temperature and nearest neighbor hopping (NNH) at low temperature.[15, 16, 19] The conductivity at the Dirac point is given by

$$\sigma_{\text{Total}} = \sigma_{\text{TA}}^0 \exp\left[-\frac{E_g}{2k_B T}\right] + \sigma_{\text{NNH}}^0 \exp\left[-\frac{E_0}{k_B T}\right], \quad (1)$$

where $\sigma_{\text{TA}}^0$ and $\sigma_{\text{NNH}}^0$ are prefactors and $E_0$ is the hopping energy. Because all the conductivity data are well fitted, $E_g$ and $E_0$ are extracted and plotted as a function of $\bar{D}$ in **Fig. 2(d)**. The maximum band gap extracted from the temperature dependence in the all-2D heterostructure was ~90 meV at $\bar{D}$ = -1.34 V/nm, which is limited due to the small dielectric constant of $h$-BN and the dielectric breakdown field of ~1.5 V/nm for the thickness used in this experiment.[48] The measured $E_g$ was smaller than the value obtained from tight-binding calculations[7] and density functional theory[49] for BLG in vacuum. However, there has been a report predicting that the band gap opened at the same $\bar{D}$ in BLG encapsulated by $h$-BN will be smaller than that in BLG in vacuum due to the screening of the electrical field by $h$-BN.[50] Due to NNH occurring by carrier hopping between localized states in the band gap, $E_0$ can be used as a measure for the amount of localized states in the band gap; i.e., the larger $E_0$ is, the smaller the amount of localized states in the band gap. In the all-2D heterostructure BLG device, $E_0$ was extracted to be 14 meV at the maximum $\bar{D}$ = -1.34 V/nm, as shown in the inset of **Fig. 2(d)**. This is much larger than $E_0$ = 2.8 meV at a larger value of $\bar{D}$ = -3.1 V/nm in the high-$k$ Y$_2$O$_3$/BLG device,[19] supporting that there are fewer impurity-induced localized states in the band gap.

**Carrier response in energy gap and DOS determination.**

The drastic reduction of charged impurities expected from the large $I_{\text{on}}/I_{\text{off}}$ also suggests the reduction of the mobile carrier response at gap states, which will be revealed here. To measure the capacitance of the limited active channel area in the all-2D heterostructure, it is critical to reduce parasitic capacitance ($C_{\text{Para}}$). In this experiment, the total capacitance between the top gate and source/drain was measured, therefore the observed $C_{\text{Para}}$ is that between the top gate and source/drain electrodes. The detailed information on the $C$-$V$ measurement is described in **Note 2**. When devices were fabricated on $n^+$-Si wafer, a large frequency dependence in $C_{\text{Para}}$ was observed because of SiO$_2$ behaving as a capacitor between the metal electrode pad and the $n^+$-Si wafer, as shown in **Fig. S7**. On the other hand, when devices were fabricated on a quartz substrate, the frequency dependence in $C_{\text{Para}}$ disappeared, resulting in negligible $C_{\text{Para}}$. Therefore, all the data shown below are obtained from the all-2D heterostructure BLG on a quartz substrate.

The total capacitance ($C_{\text{Total}}$) between the top gate and source was measured at a frequency of 1 MHz by sweeping $V_{\text{TG}}$ at each fixed $V_{\text{BG}}$ in vacuum at 20 K, as shown in **Fig. 3(a)**. Constant $\bar{D}$ lines, which indicate the constant band structure, are depicted by black dotted lines. $C_{\text{Total}}$ is composed of the top gate capacitance ($C_{\text{TG}}$) and the quantum capacitance ($C_{\text{Q}}$) of BLG and $C_{\text{Para}}$, as shown by the simplified equivalent circuit in **Fig. 3(b)**. $C_{\text{Q}}$ is regarded as the energy cost of inducing carriers in BLG and is directly given as $C_{\text{Q}} = q^2 DOS$, where $q$ and $DOS$ are the elementary charge and the density of states, respectively.[51] When the band gap is formed, $C_{\text{Q}}$ at the Dirac point decreases due to the drastic reduction of $DOS$, resulting in the reduction of $C_{\text{Total}}$. First, the capture and emission process of mobile carriers at gap states is discussed based on the frequency dependence of $C_{\text{Total}}$. The frequency dependence of $C_{\text{Total}}$ measured along the constant $\bar{D}$ = -1.25 V/nm in a different device is shown in **Fig. 3(c, d)**. No frequency dependence of $C_{\text{Total}}$ was observed in the frequency range of 10 kHz to 2 MHz. Even by the conductance method,[1] in which the capture and emission of mobile carriers by trap levels in the band gap can be detected as a deviation from the ideal carrier response, no detectable deviation from the ideal response was found, suggesting a negligible $D_{\text{it}}$ in the present frequency range. On the other hand, for all-2D heterostructure devices with bubbles in the BLG channel, a frequency dependence in $C_{\text{Total}}$ was clearly observed, as shown in **Fig. S8**. Moreover, in parallel conductance extracted from the equivalent circuit, the left shoulder of conductance peaks was observed in the high frequency region as shown in **Fig. S8(e)**, suggesting that the time constant is faster than the measurable range in this experiment. This is probably due to the present band gap opened using h-BN as insulators being small compared to the maximum band gap of ~300 meV in high-$k$ $Y_2O_3$/BLG device.[19] The clear observation of the carrier response in devices with bubbles in the BLG channel validates the interpretation of the absence of a frequency response in **Fig. 3(c, d)**.

Next, the *DOS* of BLG in all-2D heterostructure device was derived from $C_Q$ extracted from the measurement. Because of the negligible $D_{it}$, the contribution of interface trap capacitance ($C_{it}$) to $C_{Total}$ can be neglected. Therefore, by taking the high frequency limit, $C_Q$ is expressed as the simplified equivalent circuit in **Fig. 3(b)**. $C_{TG}$ was estimated to be 0.287 µF/cm² using the dielectric constant of 4 for h-BN with a thickness of 12.3 nm measured by AFM, which is roughly consistent with the h-BN capacitance at a different top gate position shown by a red arrow in **Fig. 1(b)**. Then, $C_Q$ was extracted at each $\bar{D}$ as a function of Fermi energy ($E_F$) by measuring $C_{Total}$ along the constant $\bar{D}$ using $C_{Para}$ as a fitting parameter, as shown in **Fig. 4(a)**. Although devices fabricated on a quartz substrate usually shows extremely small $C_{Para}$ as shown in **Fig. S7**, $C_{Para}$ = 0.003361 µF/cm², which is equivalent to a capacitance of 1 fF measured in the top gate area of 29.75 µm², was used to produce better fits. The extracted $C_Q$ at $\bar{D}$ = 0 V/nm fits reasonably well with the theoretical value for bilayer graphene calculated by the tight-binding model.[52] $E_F$ was calculated by the same method as in previously reported measurement.[19] $E_F$ is expressed as $E_F$ =

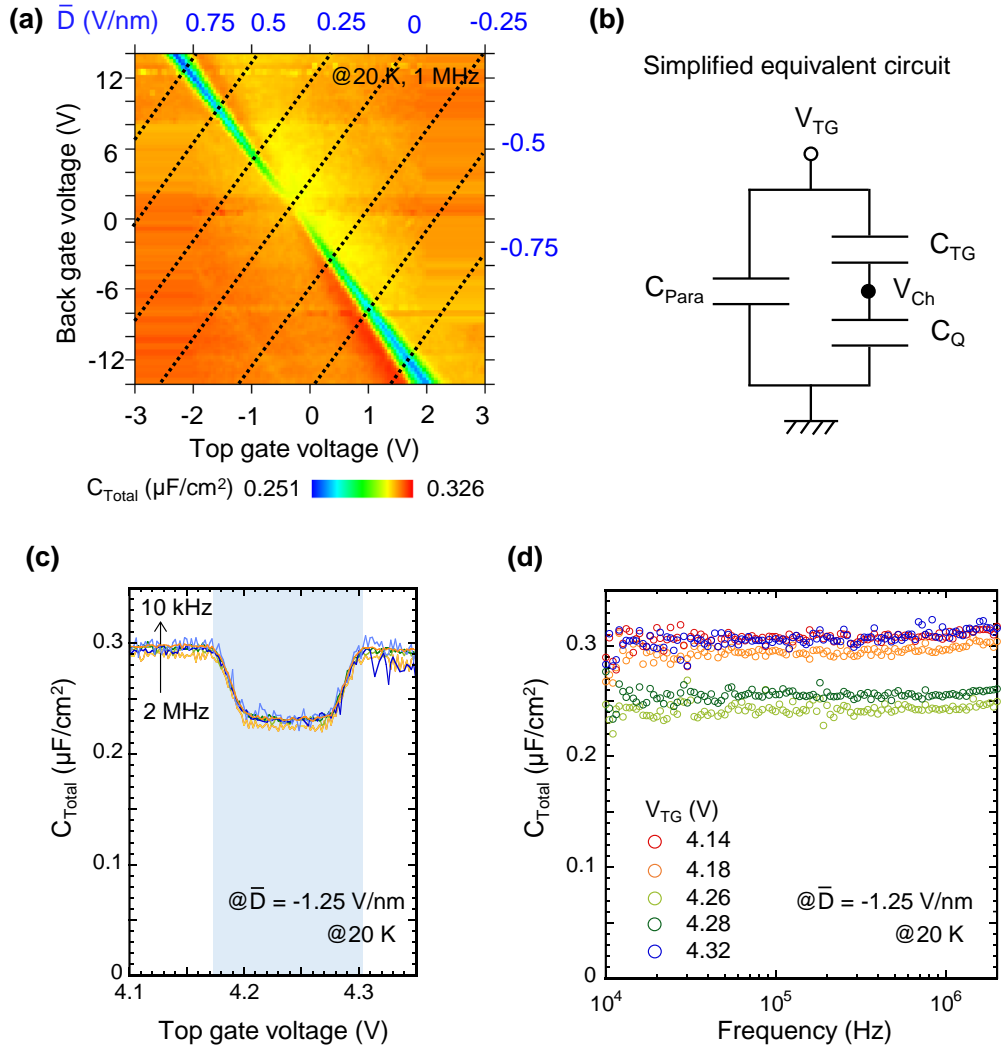

**Figure 3** (a) Color plot of $C_{Total}$ measured at a frequency of 1 MHz at 20 K. (b) Simplified equivalent circuit for *C-V* measurement in BLG-FET. (c) $C_{Total}$ as a function of $V_{TG}$ measured at constant $\bar{D}$ = -1.25 V/nm in a different device at 20 K. It should be noted that $V_{BG}$ is also changed to maintain constant $\bar{D}$. (d) $C_{Total}$ as a function of frequency measured at each $V_{TG}$ in (c).

$qV_{Ch}$ which can be calculated from a relationship of serial capacitors according to $V_{Ch} = V'_{TG} - \int_0^{V_{TG}} C'_{Total}/C_{TG} dV'_{TG}$. $V_{TG}'$ and $C_{Total}'$ are defined as $V_{TG}' = V_{TG} - V_{CN}$, where $V_{CN}$ is the charge neutrality voltage for each $V_{TG}$, and $C_{Total}' = C_{Total} - C_{Para}$, respectively. As $\bar{D}$ is increased, $C_Q$ at the Dirac point decreases due to band gap opening in BLG. In gapped BLG, van Hove singularities (vHS) are clearly observed at gap edges. Increasing disorder, such as charged impurities, results in more smearing of vHS and more disorder-induced states in the band gap, leading to gap closing at small $\bar{D}$.[19, 53, 54] The measured result shows two sharp vHS at both gap edges and a well-defined gap, indicating that the disorder strength is considerably reduced compared to the previous experiments on h-BN/BLG on SiO$_2$/Si and high-$k$ Y$_2$O$_3$/BLG on SiO$_2$/Si devices where only one pronounced vHS was observed.[19, 54] The vHS observed were asymmetrical, which can be explained by the near-layer capacitance enhancement effect.[55] Here, the $C_Q$ at the Dirac point ($C_Q$@DP) can be used as a measure of the amount of disorder-induced states in the band gap. The $C_Q$@DP values for both all-2D heterostructure and high-$k$ Y$_2$O$_3$/BLG devices are shown in **Fig. 4(b)**. The result shows that in the all-2D heterostructure device, $C_Q$@DP decreases rapidly even at small $\bar{D}$ and almost reaches zero compared to that in the high-$k$ Y$_2$O$_3$/BLG device,[19] supporting that disorder was reduced in the all-2D heterostructure. The other study also found that the graphite top gate could further improve the resolution on the integer quantum Hall states.[45] Since vHS are clearly observed, the size of the band gap opened at each $\bar{D}$ was evaluated by defining the band gap size as the energy difference between vHS at both sides. As shown in **Fig. 2(d)**, the measured results from $C$-$V$ and $I$-$V$ in all-2D heterostructure devices are consistent even at low $\bar{D}$.

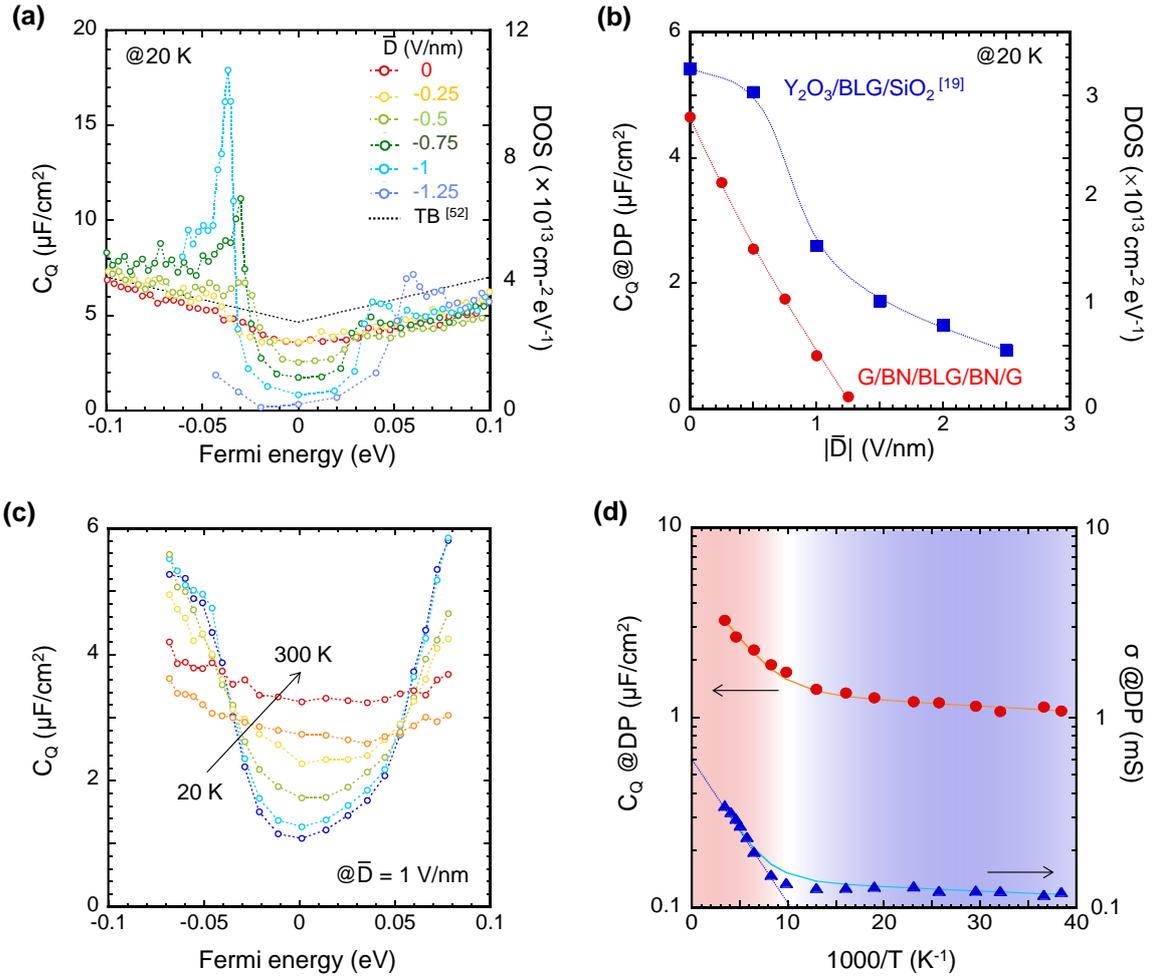

**Figure 4** (a) $C_Q$ as a function of Fermi energy at $\bar{D}$ = 0 to -1.25 V/nm at 20 K. (b) $C_Q$ at the Dirac point in (a) as a function of $|\bar{D}|$. (c) $C_Q$ measured at $|\bar{D}|$ = 1 V/nm at temperatures from 20 K to 300 K for the different sample as in (a). (d) $C_Q$ at the Dirac point in (c) and $\sigma$ at the Dirac point of the same device in (c) at the same $|\bar{D}|$ as a function of inverse temperature. Solid lines are fits.

To investigate the behavior of disorder-induced states in the band gap, the temperature dependence of $C_Q$ in the band gap of the all-2D heterostructure was observed using a different device as in **Fig. 4(a)**. **Fig. 4(c)** shows $C_Q$ as a function of $E_F$ measured at $\bar{D}$ = 1 V/nm at each temperature. At finite temperature, the $C_Q$ of a 2D material is given by

$$C_Q = q^2 g_{2D}\left[1 + \frac{\exp(E_g/2k_B T)}{2\cosh(qV_{Ch}/k_B T)}\right]^{-1}, \quad (2)$$

where $g_{2D} = g_s g_v m^*/2\pi\hbar^2$ is the band-edge *DOS*, which was chosen from the experimental value, and $k_B$ is the Boltzmann constant.[56] $g_s$ and $g_v$ are the spin and valley degeneracy factors, respectively, and $m^*$ is the effective mass. This results in the temperature dependence of $C_Q$ in the band gap. $C_Q$@DP is plotted as a function of temperature in **Fig. 4(d)**. Only the data in the high temperature range could be fitted by eq. (2). The temperature dependence of $C_Q$@DP measured suggests a different mechanism at the low temperature range. The transition in the dominant mechanism in $C_Q$@DP seemed to be in the same temperature range as σ@DP. It was found that adding the NNH term using $E_0$ measured at the same $\bar{D}$ to eq. (1) resulted in all the calculated values being fitting with the measured data well at all temperatures, as shown in **Fig. 4(d)**. This indicates that although there was no frequency response of capacitance in the band gap, quite a small amount of disorder-induced states still exists in the band gap. It may be possible that these remnant gap states have a time constant even faster than those observed in **Fig. S8(e)** because the maximum band gap opened is small.

## Quantitative estimation of the potential fluctuation

The reduction of charged impurities has been qualitatively demonstrated from both the *I-V* and *C-V* analyses. Here, let us extract the potential fluctuation due to charged impurities in the energy unit quantitatively. Ideally, $E_g$ should be zero at $\bar{D}$ = 0 V/nm, but in experiments, finite $E_g$ at $\bar{D} \sim 0$ V/nm has been observed in previous studies because the potential fluctuation prevents the gap from closing locally. Therefore, remnant $E_g/2$ value at $\bar{D} \sim 0$ V/nm is associated with the root-mean-square amplitude of the potential fluctuation ($\Phi_{rms}$) in BLG.[15] **Figure 5(a)** compares $E_g/2$ as a function of $\bar{D}$ for the all-2D heterostructure and high-*k* Y$_2$O$_3$/BLG device. In our previous high-*k* Y$_2$O$_3$/BLG device,[19] $\Phi_{rms}$ was estimated to be 26 meV which is in agreement with the previous literature data on other high-*k*/BLG cases.[15] A significantly smaller value of $\Phi_{rms}$ = ~0.5 meV was observed in several all-2D heterostructure devices fabricated in this study. The carrier density equivalent to this potential fluctuation was estimated to be 1.3×10$^{10}$ cm$^{-2}$ by the relation of $n = \Phi_{rms}(2m^*/\pi\hbar^2)$, where a constant 2D *DOS* for small energy deviation of $E_F$ is assumed for simplicity. It should be emphasized that BLG is most sensitive to the potential fluctuation because the carrier density difference between the top and bottom graphene layers results in band gap opening, suggesting that the utilization of all-2D heterostructure is critical. Moreover, although the observed potential fluctuation in the all-2D heterostructure seems to be comparable to those extracted from recently reported high-quality *h*-BN-

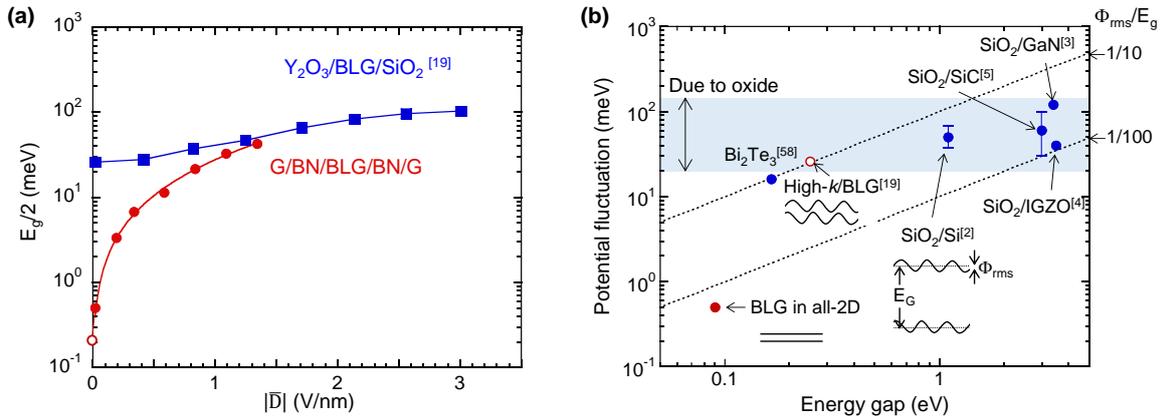

**Figure 5** (a) $E_g/2$ as a function of $\bar{D}$ for the all-2D heterostructure and high-*k* Y$_2$O$_3$/BLG devices. The remnant $E_g/2$ value at $\bar{D} \sim 0$ V/nm can be regarded as $\Phi_{rms}$ in BLG. (b) Comparison of potential fluctuations in various insulator/semiconductor gate stack structures. Schematic illustration on spatial fluctuation of $E_G$ and $\Phi_{rms}$ for SiO$_2$/Si, high-*k*/BLG, and BLG in all-2D are shown in the figure.

encapsulated BLG on $SiO_2$/Si[34,36] or on the metal electrode,[57] the difference between with and without the graphite back gate electrode becomes more obvious in the $I_{on}$/$I_{off}$ data in **Fig. 2(b)**. This indicates that the role of the graphite back and top gate electrodes is to apply uniform $\bar{D}$ as well as to screen charged impurities in $SiO_2$.

**Figure 5(b)** summarizes the value of the band gap and $\Phi_{rms}$ due to charged impurities and fixed charges in various semiconductor/insulator systems. Although these values have been evaluated in different ways, a comparison is roughly possible because of the same physical meaning. The hatched region (20~100 meV) indicates the potential fluctuation induced inevitably due to the amorphous oxide insulator, while the dotted and dot-dashed lines indicates $\Phi_{rms}/E_g$ = 1/10 and 1/100, respectively. The effect of the potential fluctuation due to amorphous oxide insulators is more prominent in small-gap semiconductors and topological insulators.[58] Therefore, the van der Waals layered heterostructure is quite promising to study the physical properties with the small energy (a few meV).

**DISCUSSION**

To realize BLG-FETs, that is, to achieve sufficient $I_{on}$/$I_{off}$ at room temperature, it is necessary to open the maximum band gap of ~0.3 eV while retaining the present BLG/$h$-BN interface quality. This requires applying $|\bar{D}| \geq 3$ V/nm, which is not attainable only with $h$-BN as insulators due to its small dielectric constant of ~4. Although the combination of high-$k$ amorphous oxide and $h$-BN provides $|\bar{D}| \geq 3$ V/nm, the potential fluctuation due to charged impurities in high-$k$ amorphous oxide will deteriorate $I_{on}$/$I_{off}$. To apply $|\bar{D}| \geq 3$ V/nm, the gate stack structure needs to be reconsidered, such as by combining $h$-BN with high-$k$ single-crystal nanosheets,[59, 60] in which potential fluctuation might be reduced due to crystal periodicity.

Finally, let us discuss the resistivity saturation at DP with increasing $\bar{D}$ observed often in the previous literatures.[36, 61] Two different origins are proposed. One is the topological current along the AB-BA domain boundary in gapped BLG and the resistivity saturation has been observed at around $\bar{D}$ ~1 V/nm.[61] In this study, such behavior was not observed as the resistivity continued to decrease up to $\bar{D}$ = -1.48 V/nm. This $\bar{D}$ value is one of the best values for $h$-BN encapsulated BLG so far and is sufficient to discuss the conductivity saturation. Therefore, it is suggested that there was no contribution of the current along the AB-BA domain boundary. The other origin is the topological edge current in gapped BLG, as mentioned in Introduction. The device with the edgeless-Corbino geometry showed no resistivity saturation within the maximum $\bar{D}$ value of ~0.6 V/nm, while the resistivity of the device with the Hall bar geometry saturated at around $\bar{D}$ = ~0.2 V/nm.[36] This contradicts the present study. In general, the edge current requires the zigzag edge structure, and the channel length should be shorter than the localization length. In this study, the channel length of ~20 μm could be much longer than the localization length, resulting in the high $I_{on}$/$I_{off}$. However, for a short channel, the edge structure should also be controlled to prevent the topological edge current.

**CONCLUSIONS**

We have demonstrated significantly improved device properties in all-2D heterostructure BLG-FETs on quartz compared to $h$-BN-encapsulated BLG on $SiO_2$/Si as well as high-$k$ $Y_2O_3$/BLG on $SiO_2$. For the transfer characteristics at 20 K, $I_{on}$/$I_{off}$ reached 4.6×10⁵ with the off-state resistivity of ~5GΩ and the field effect mobility was 20000 cm²/Vs. For the $h$-BN/BLG interface quality, the non-detectable amount of $D_{it}$ and two sharp vHS suggests that the 2D heterointerface is electrically inert. These significant results are mainly attributed to the drastic reduction of the potential fluctuation to ~1 meV and hence the spatial uniformity of $E_G$. This quite low potential fluctuation can be achieved only in all-2D heterostructure BLG-FETs, not in conventional semiconductor systems with high-$k$ gate stacks. Therefore, the all-2D heterosturcture BLG-FET is promising for future electronics applications.

**SUPPORTING INFORMATION**

Raman spectra of BLG, bubble free transfer, electrode formation to encapsulated heterostructure, EBSD analysis on top gate metal electrode, contribution of Parasitic capacitance to the total capacitance, and Frequency dispersion for the heterostructure with bubbles. The Supporting Information is available free of charge via the Internet at http://pubs.acs.org.

**AUTHOR INFORMATION**
**Corresponding Author**
*Email: nagashio@material.t.u-tokyo.ac.jp
**Notes**
The authors declare no competing financial interests.

**ACKNOWLEDGEMENTS**
We are grateful to Covalent Materials for kindly providing us with Kish graphite. This research was supported by the JSPS Core-to-Core Program, A. Advanced Research Networks, JSPS KAKENHI Grant Numbers JP25107004, JP16H04343, JP16K14446, and JP26886003, and JST PRESTO Grant Number JPMJPR1425, Japan.

**REFERENCES**

[1] Nicolian, E. H.; Brews, J. R. MOS Physics and Technology. Wiley, New York, 1982.

[2] Declerck, G.; Van Overstraeten, R.; Broux, G. Discussion of The Surface-Potential Fluctuations Caused by Oxide-Charge Fluctuations. *J. Appl. Phys.* **1974**, *45*, 2593.

[3] Gaffey, B.; Guido, L. J.; Wang, X. W.; Ma, T. P. High-Quality Oxide/Nitride/Oxide Gate Insulator for GaN MIS structures. *IEEE T. Elec. Dev.* **2001**, *48*, 458-464.

[4] Lee, S.; Ghaffarzadeh, K.; Nathan, A.; Robertson, J.; Jeon, S.; Kim, C.; Song, I.; Chung, U. Trap-Limited and Percolation Conduction Mechanisms in Amorphous Oxide Semiconductor Thin Film Transistors. *Appl. Phys. Lett.* **2011**, *98*, 203508.

[5] Bano, E.; Ouisse, T.; Cioccio, L. D.; Karmann, S. Surface Potential Fluctuations in Metal-Oxide-Semiconductors Capacitors Fabricated on Different Silicon Carbide Polytypes. *Appl. Phys. Lett.* **1994**, *65*, 2723.

[6] Displacement field ( $\bar{D}$ ) is defined by $\bar{D} = 1/2[\varepsilon_{BG}/d_{BG}(V_{BG} - V^0_{BG}) - \varepsilon_{TG}/d_{TG}(V_{TG} - V^0_{TG})]$, where $\varepsilon_{BG}$, $\varepsilon_{BG}$, $d_{BG}$, $d_{TG}$, $V_{BG}$, $V_{TG}$, are dielectric constants, insulator thicknesses, gate voltages of the back and top gate insulators, respectively. $V^0_{BG}$ and $V^0_{TG}$ are the charge neutrality point in the top gated region. The dual gated bilayer device allows independent control of the carrier density and the size of the band gap by applying $V_{BG}$ and $V_{TG}$.

[7] Zhang, Y.; Tang, T.; Girit, C.; Hao, Z.; Martin, M. C.; Zettl, A.; Crommie, M. F.; Shen, Y. R.; Wang, F. Direct Observation of A Widely Tunable Bandgap in Bilayer Graphene. *Nature* **2009**, *459*, 820-823.

[8] McCann, E.; Fal'ko, V. I. Landau-Level Degeneracy and Quantum Hall Effect in A Graphite Bilayer. *Phys. Rev. Lett.* **2006**, *96*, 086805.

[9] Zhu, W.; Perebeinos, V.; Freitag, M.; Avouris, P. Carrier Scattering, Mobilities, and Electrostatic Potential in Monolayer, Bilayer, and Trilayer Graphene. *Phys. Rev. B* **2009**, *80*, 235402.

[10] Deshpande, A.; Bao, W.; Zhao, Z.; Lau, C. N.; Leroy, B. J. Mapping the Dirac Point in Gated Bilayer Graphene. *Appl. Phys. Lett.* **2009**, *95*, 243502.

[11] Oostinga, J. B.; Heersche, H. B.; Liu, X.; Morpurgo, A. F.; Vandersypen, L. M. K. Gate-Induced Insulating State in Bilayer Graphene Devices. *Nat. Mater.* **2008**, *7*, 151-157.

[12] Xia, F.; Farmer, D. B.; Lin, Y.; Avouris, P. Graphene Field-Effect Transistors with High On/Off Current Ratio and Large Transport Band Gap at Room Temperature. *Nano Lett.* **2010**, *10*, 715-718.

[13] Miyazaki, H.; Tsukagoshi, K.; Kanda, A.; Otani, M.; Okada, S. Influence of Disorder on Conductance in Bilayer Graphene under Perpendicular Electric Field. *Nano Lett.* **2010**, *10*, 3888-3892.

[14] Yan, J.; Fuhrer, M. S. Charge Transport in Dual Gated Bilayer Graphene with Corbino Geometry. *Nano Lett.* **2010**, *10*, 4521-4525.

[15] Zou, K.; Zhu, J. Transport in Gapped Bilayer Graphene: The Role of Potential Fluctuations. *Phys. Rev. B* **2010**, *82*, 081407(R).

[16] Taychatanapat, T.; Jarillo-Herrero, P. Electronic Transport in Dual-Gated Bilayer Graphene at Large Displacement Fields. *Phys. Rev. Lett.* **2010**, *105*, 166601.

[17] Szafranek, B. N.; Fiori, G.; Schall, D.; Neumaier, D.; Kurz, H. Current Saturation and Voltage Gain in Bilayer Graphene Field Effect Transistors. *Nano Lett.* **2012**, *12*, 1324-1328.

[18] Lee, K.; Fallahazad, B.; Min, H.; Tutuc, E. Transport Gap in Dual-Gated Graphene Bilayers Using Oxides as Dielectrics. *IEEE T. Elecron Dev.* **2013**, *60*, 103-108.

[19] Kanayama, K.; Nagashio, K. Gap State Analysis in Electric Field-Induced Band Gap for Bilayer Graphene. *Sci. Rep.* **2015**, *5*, 15789.

[20] Chu, T.; Chen, Z. Achieving Large Transport Bandgaps in Bilayer Graphene. *Nano Res.* **2015**, *8*, 3228-3236.

[21] Dean, C. R.; Young, A. F.; Meric, I.; Lee, C.; Wang, L.; Sorgenfrei, S.; Watanabe, K.; Taniguchi, T.; Kim, P.; Shepard, K. L.; Hone, J.; Boron Nitride Substrates for High-Quality Graphene Electronics. *Nat. Nanotechnol.* **2010**, *5*, 722-726.

[22] Wang, H.; Taychatanapat, T.; Hsu, A.; Watanabe, K.; Taniguchi, T.; Jarillo-Herrero, P.; Palacious, T. BN/Graphene/BN Transistors for RF Applications. *IEEE Electron Device Lett.* **2011**, *32*, 1209-1211.

[23] Goossens, A.; Driessen, A. C. M.; Baart, T. A.; Watanabe, K.; Taniguchi, T.; Vandersypen, M. K. Gate-Defined Confinement in Bilayer Graphene-Hexagonal Boron Nitride Hybrid Devices. *Nano Lett.* **2012**, *12*, 4656-4660.

[24] Dean, C. R.; Wang, L.; Maher, P.; Forsythe, C.; Ghahari, F.; Gao, Y.; Katoch, J.; Ishigami, M.; Moon, P.; Koshino, M.; Taniguchi, T.; Watanabe, K.; Shepard, K. L.; Hone, J.; Kim, P. Hofstadter's Butterfly and The Fractal Quantum Hall Effect in Moire Superlattices. *Nature* **2013**, *497*, 598-602.

[25] Varlet, A.; Bischoff, D.; Simonet, P.; Watanabe, K.; Taniguchi, T.; Ihn, T.; Ensslin, K.; Mucha-Kruczynski, M.; Fal'ko, V. I. Anomalous Sequence of Quantum Hall Liquids Revealing a Tunable Lifshitz Transition in Bilayer Graphene. *Phys. Rev. Lett.* **2014**, *113*, 116602.

[26] Shimazaki, Y.; Yamamoto, M.; Borzenets, I. V.; Watanabe, K.; Taniguchi, T.; Tarucha, S. Generation and Detection of Pure Valley Current by Electrically Induced Berry Curvature in Bilayer Graphene. *Nat. Phys.* **2015**, *11*, 1032-1036.

[27] Sui, M.; Chen, G.; Ma, L.; Shan, W.; Tian, D.; Watanabe, K.; Taniguchi, T.; Jin, X.; Yao, W.; Xiao, D.; Zhang, Y. Gate-Tunable Topological Valley Transport in Bilayer Graphene. *Nat. Phys.* **2015**, *11*, 1027-1031.

[28] Kumar, C.; Kuiri, M.; Jung, J.; Das, T.; Das, A. Tunability of 1/f Noise at Multiple Dirac Cones in hBN Encapsulated Graphene Devices. *Nano Lett.* **2016**, *16*, 1042-1049.

[29] Avsar, A.; Vera-Marun, I. J.; Tan, J.; Koon, G.; Watanabe, K.; Taniguchi, T.; Shaffique, A.; Ozyilmaz, B. Electronic Spin Transport in Dual-Gated Bilayer Graphene. *NPG Asia Mater.* **2016**, *8*, e274.

[30] Singh, S.; Katouch, J.; Zhu, T.; Meng, K.; Liu, T.; Brangham, J. T.; Yang, F.; Flatte, M. E.; Kawakami, R. K. Strong Modulation of Spin Currents in Bilayer

Graphene by Static and Fluctuating Proximity Exchange Fields. *Phys. Rev. Lett.* **2017**, *118*, 187201.

[31] Kim, K.; Yankowitz, M.; Fallahazad, B.; Kang, S.; Movva, H. C. P.; Huang, S.; Larentis, S.; Corbet, C. M.; Taniguchi, T.; Watanabe, K.; Banerjee, S. K.; Leroy, B. J.; Tutuc, E. van der Waals Heterostructures with High Accuracy Rotational Alignment. *Nano lett.* **2016**, *16*, 1989-1995.

[32] Nguyen, V. L.; Perello, D. J.; Lee, S.; Nai, C. T.; Shin, B. G.; Kim, J.; Park, H. Y.; Jeong, H. Y.; Zhao, J.; Vu, Q. A.; Lee, S. H.; Loh, K. P.; Jeong, S.; Lee, Y. H. Wafer-Scale Single-Crystalline AB-Stacked Bilayer Graphene. *Adv. Mater.* **2016**, *28*, 1-7.

[33] Hao, Y.; Wang, L.; Liu, Y.; Chen, H.; Wang, X.; Tan, T.; Nie, S.; Suk, J. W.; Jiang, T.; Liang, T.; Xiao, J.; Ye, W.; Dean, C. R.; Yakobson, B. I.; McCarty, K. F.; Kim, P.; Hone, J.; Colombo, L.; Ruoff, R. S. Oxygen-Activated Growth and Bandgap Tunability of Large Single-Crystal Bilayer Graphene. *Nat. Nanotechnol.* **2016**, 11, 426-431.

[34] Schmitz, M.; Engels, S.; Banszerus, L.; Watanabe, K.; Taniguchi, T.; Stampfer, C.; Beschoten, B. High Mobility Dry-Transferred CVD Bilayer Graphene. *Appl. Phys. Lett.* **2017**, *110*, 263110.

[35] Li, J.; Wang, K.; McFaul, K. J.; Zern, Z.; Ren, Y.; Watanabe, K.; Taniguchi, T.; Qiao, Z.; Zhu, J. Gate-Controlled Topological Conducting Channels in Bilayer Graphene. *Nat. Nanotechnol.* **2016**, *11*, 1060-1065.

[36] Zhu, M. J.; Kretinin, A. V.; Thompson, M. D.; Bandurin, D. A.; Hu, S.; Yu, G. L.; Birkbeck, J.; Mishchenko, A.; Vera-Marun, I. J.; Watanabe, K.; Taniguchi, T.; Polini, M.; Prance, J. R.; Novoselov, K. S.; Geim, A. K.; Shalom, M. B. Edge Currents Shunt The Insulating Bulk in Gapped Graphene. *Nat. Commun.* **2017**, *8*, 14552.

[37] Lee, J.; Watanabe, K.; Taniguchi, T.; Lee, H. Realization of Topological Zero-Energy Mode in Bilayer Graphene in Zero Magnetic Field. *Sci. Rep.* **2017**, *7*, 6466.

[38] Zomer, P. J.; Dash, S. P.; Tombros, N.; van Wees, B. J. A Transfer Technique for High Mobility Graphene Devices on Commercially Available Hexagonal Boron Nitride. *App. Phys. Lett.* **2011**, *99*, 232104.

[39] Maher, P.; Wang, L.; Gao, Y.; Forsythe, C.; Taniguchi, T.; Watanabe, K.; Abanin, D.; Papic, Z.; Cadden-Zimansky, P.; Hone, J.; Kim, P.; Dean, C. R. Tunable Fractional Quantum Hall Phases in Bilayer Graphene. *Science* **2014**, *345*, 61-64.

[40] Sanchez-Yamagishi, J. D.; Luo, J. Y.; Young, A. F.; Hunt, B. M.; Watanabe, K.; Taniguchi, T.; Ashoori, R. C.; Jarillo-Herrerro, P. Helical Edge States and Fractional Quantum Hall Effect in A Graphene Electron-Hole Bilayer. *Nat. Nanotechnol.* **2017**, *12*, 118-122.

[41] Hunt, B. M.; Li, J. I. A.; Zibrov, A. A.; Wang, L.; Taniguchi, T.; Watanabe, K.; Hone, J.; Dean, C. R.; Zaletel, M.; Ashoori, R. C.; Young, A. F. Direct Measurement of Discrete Valley and Orbital Quantum Numbers in Bilayer Graphene. *Nature Commun.* **2017**, *8*, 948.

[42] Wang, L.; Meric, I.; Huang, P. Y.; Gao, Q.; Gao, Y.; Tran, H.; Taniguchi, T.; Watanabe, K.; Campos, L.; Muller, D. A.; Guo, J.; Kim, P.; Hone, J.; Shepard, K. L.; Dean, C. R. One-Dimensional Electrical Contact to A Two-Dimensional Material. *Science* **2013**, *342*, 614-617.

[43] Pizzocchero, F.; Gammelggard, L.; Jessen, B. S.; Caridad, J. M.; Wang, L.; Hone, J.; Boggild, P.; Booth, T. J. The Hot Pick-Up Technique for Batch Assembly of van der Waals Heterostructures. *Nat. Commun.* **2016**, *7*, 11894.

[44] Uwanno, T.; Hattori, Y.; Taniguchi, T.; Watanabe, K.; Nagashio, K. Fully Dry PMMA Transfer of Graphene on *h*-BN Using A Heating/Cooling System. *2D Mater* **2015**, *2*, 041002.

[45] Zibrov, A. A.; Kometter, C.; Zhou, H.; Spanton, E. M.; Taniguchi, T.; Watanabe, K.; Zaletel, M. P.; Young, A. F. Tunable Interacting Composite Fermion Phases in a Half-Filled Bilayer-Graphene Landau Level. *Nature* **2017**, *549*, 360-364.

[46] Overweg, H.; Eggimann, H.; Chen, X.; Slizovskiy, S.; Eich, M.; Risoni, R.; Lee, Y.; Rickhaus, P.; Watanabe, K.; Taniguchi, T.; Fal'ko, V.; Ihn, T.; Ensslin, K. Electrostatically Induced Quantum Point Contacts in Bilayer Graphene. *Nano Lett.* **2018**, *18*, 553-339.

[47] Sagade, A. A.; Neumaier, D.; Schall, D.; Otto, M.; Pesquera, A.; Centeno, A.; Elorza, A. Z.; Kurz, H. Highly Air Stable Passivation of Graphene Based Field Effect Devices. *Nanoscale* **2015**, *7*, 3558.

[48] Hattori, Y.; Taniguchi, T.; Watanabe, K.; Nagashio, K. Anisotropic Dielectric Breakdown Strength of Single Crystal Hexagonal Boron Nitride. *ACS Appl. Mater, Interfaces* **2016**, *8*, 27877-27884.

[49] Min, H.; Sahu, B.; Banerjee, S. K.; MacDonald, A. H. *Ab initio* Theory of Gate Induced Gaps in Graphene Bilayers. *Phys. Rev. B*, **2007**, *75*, 155115.

[50] Ramasubramanium, A.; Naveh, D.; Towe, E. Tunable Band Gaps in Bilayer Graphene-BN Heterostructures. *Nano Lett.* **2011**, *11*, 1070-1075.

[51] Fang, T.; Konar, A.; Xing, H.; Jena, D. Carrier Statistics and Quantum Capacitance of Graphene Sheets and Ribbons. *Appl. Phys. Lett.* **2007**, *91*, 092109.

[52] Koshino, M. Electronic Transport in Bilayer Graphene. *New J. Phys.* **2009**, *11*, 095010.

[53] Nilsson, J.; Castro Neto, A. H. Impurities in A Biased Graphene Bilayer. *Phys. Rev. Lett.* **2007**, *98*, 126801.

[54] Young, A. F.; Dean, C. R.; Meric, I.; Sorgenfrei, S.; Ren, H.; Watanabe, K.; Taniguchi, T.; Hone, J.; Shepard, K. L.; Kim, P. Electronic Compressibility of Layer-Polarized Bilayer Graphene. *Phys. Rev. B* **2012**, *85*, 235458.

[55] Young, A. F.; Levitov, L. S. Capacitance of Graphene Bilayer as A Probe of Layer-Specific Properties. *Phys. Rev. B* **2011**, *84*, 085441.

[56] Ma, N.; Jena, D. Carrier Statistics and Quantum Capacitance Effects on Mobility Extraction in Two-Dimensional Crystal Semiconductor Field-Effect Transistors. *2D Mater.* **2015**, *2*, 015003.

[57] Cao, Y.; Fatemi, V.; Fang, S.; Watanabe, K.; Taniguchi, T.; Kaxiras, E.; Jarillo-Herrero, P. Unconventional Superconductivity in Magic-Angle Graphene Superlattices. *Nature* **2018**, *556*, 43-50.

[58] Beidenkopf, H.; Roushan, P.; Seo, J.; Gorman, L.;

Drozdov, I.; Hor, Y.; Cava, R. J.; Yazdani, A. Spatial Fluctuations of Helical Dirac Fermions on The Surface of Topological Insulators. *Nat. Phys.* **2011**, *7*, 939-943.

[59] Osada, M.; Akatsuka, K.; Ebina, Y.; Funakubo, H.; Ono, K.; Takada, K.; Sasaki, T. Robust High-*k* Response in Molecularly Thin Perovskite Nanosheets. *ACS Nano* **2010**, *4*, 5225-5232.

[60] Sekizaki, S.; Osada, M.; Nagashio, K. Molecularly-Thin Anatase Field-Effect Transistors Fabricated Through The Solid State Transformation of Titania Nanosheets. *Nanoscale* **2017**, *19*, 6471-6477.

[61] Ju, L.; Shi, Z.; Nair, N.; Lv, Y.; Jin, C.; Velasco, J.; Ojerda-Aristizabal, C.; Bechtel, H. A.; Martin, M. C.; Zettl, A.; Analytis, J.; Wang, F. Topological Valley Transport at Bilayer Graphene Domain Walls. *Nature* **2015**, *520*, 650-655.

# Electrically inert *h*-BN/bilayer graphene interface in all-2D-heterostructure FETs

*Teerayut Uwanno,*[†,‡] *Takashi Taniguchi,*[§] *Kenji Watanabe,*[§] *and Kosuke Nagashio\**[†,∥]

[†]Department of Materials Engineering, The University of Tokyo, Tokyo 113-8656, Japan

[‡]College of Nanotechnology, King Mongkut's Institute of Technology Ladkrabang, Bangkok 10520, Thailand

[§]National Institute of Materials Science, Ibaraki 305-0044, Japan

[∥]PRESTO, Japan Science and Technology Agency (JST), Tokyo 113-8656, Japan

**E-mail:** nagashio@material.t.u-tokyo.ac.jp

**Note 1. Details of device fabrication**

All-2D heterostructures were fabricated by a method similar to previously reported pick up method using polymers of polydimethylsiloxane (PDMS) and polymethylmethacrylate (PMMA). BLG, *h*-BN and graphite were mechanically exfoliated onto each $SiO_2$ (90 nm)/$n^+$-Si wafer, while *h*-BN for the top gate insulator and graphite for the top gate electrode were separately exfoliated onto PMMA/PDMS/glass slide substrates whose surfaces were treated by oxygen plasma. Using a micromanipulator system, *h*-BN top gate and BLG were aligned and then heated to ~100°C for 5 minutes to remove adsorbed water on their surfaces. Afterwards, the substrates were brought into contact while being heated to ~55°C to facilitate adhesion between the crystals. Because the van der Waals force between 2D crystals is greater than that between a 2D crystal and $SiO_2$, a 2D crystal on $SiO_2$ can be picked up by another 2D crystal on the PMMA/PDMS/glass slide. After picking up BLG by *h*-BN top gate, *h*-BN back gate and graphite back gate electrode were picked up by the same method. The change in Raman peaks of BLG was followed during this transfer process, as shown below. After the stack was transferred onto a quartz substrate by dissolving the polymer, the stack was annealed in Ar/$H_2$ gas at 200°C for 3 hours to facilitate the aggregation of bubbles and to remove polymer residue. After annealing, graphite top gate electrode was transferred onto *h*-BN top gate followed by annealing again under the same conditions.

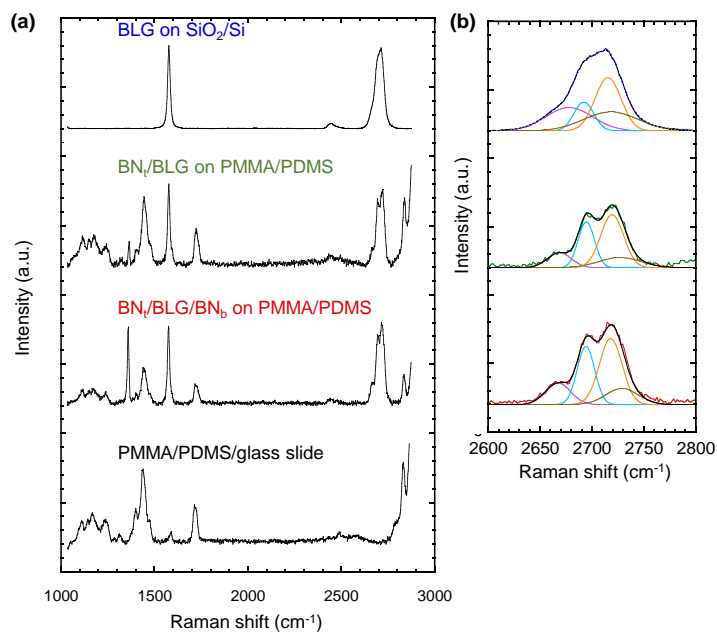

**Figure S1** (a) Normalized Raman spectra of BLG over the course of heterostructure fabrication. $BN_t$ and $BN_b$ are *h*-BN top and back gate insulators, respectively. (b) Normalized 2D peak in (a) fitted by four components. After picking up BLG with $BN_t$, the split of the components for 2D peak became pronounced because the peak width for each component became sharp.

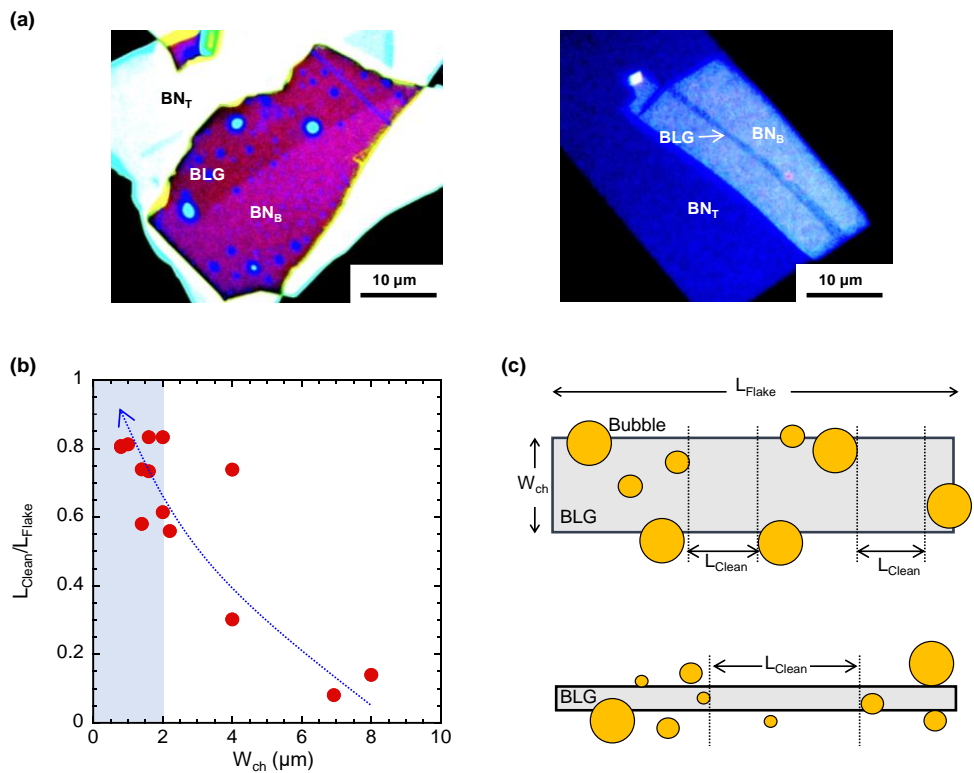

**Figure S2** (a) Optical images of *h*-BN-encapsulated BLG. Interfacial bubbles are highlighted by enhanced optical contrast. (b) Fraction of the length of clean BLG area ($L_{Clean}$) and the length of BLG flake as a function of BLG width ($W_{ch}$). (c) Definition of $L_{Clean}$ and $L_{Flake}$ in (b).

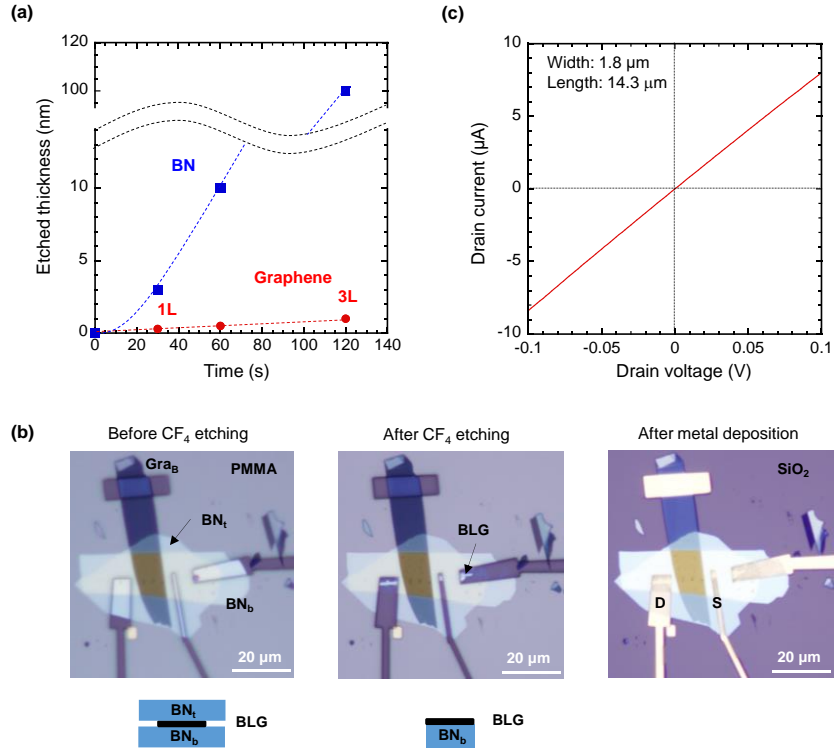

**Figure S3** (a) Etching rate as a function of etching time. (b) Optical images of electrical contact formation process. $BN_t$ and $BN_b$ is h-BN top gate and h-BN back gate respectively. After $CF_4$ etching the shape of BLG on $BN_b$ can be seen as BLG on $BN_b$ was not etched. (c) Drain current as a function of drain voltage measured between electrodes shown as S and D in (b).

In this study, h-BN was selectively etched by $CF_4$ plasma with 18 W power, exposing BLG surface for formation of planar electrical contact on BLG. The selectivity of h-BN compared to BLG is due to chemical reaction being more dominant in low power plasma, while in high power plasma physical etching is dominant. The etching rate difference between h-BN and BLG by $CF_4$ plasma shown in (a) originates from the difference between bonding energies. B-N bonding energy is 389 kJ/mol while B-F bonding energy is 613 kJ/mol [a], meaning that rather than staying as h-BN, forming B-F is more energetically favorable, leading to high reactivity of h-BN to $CF_4$ plasma. On the other hand, C-F bonding energy is 485 kJ/mol, while C=C bonding energy is 602 kJ/mol,[a] meaning that staying as graphene is more energetically favorable than forming C-F, leading to low reactivity of graphene to $CF_4$ plasma. The difference in the etching rate is high enough that, as shown in (b), 23 nm-thick h-BN top gate and 35 nm-thick h-BN back gate were completely etched while BLG remains the same, even protecting h-BN under it. After $CF_4$ etching, the presence of BLG in contact areas were confirmed by Raman spectroscopy. This enables the formation of surface electrical contact on BLG. Then, the contact areas were exposed to $O_3$ plasma in order to remove contaminants,[b] and lastly Ni/Au electrodes were deposited by thermal evaporation. The contacts formed were Ohmic contact as shown in (c).

[a] Haynes, W. M. *CRC Handbook of Chemistry and Physics 92$^{nd}$ Ed.* CRC Press, Boca Raton 2011.
[b] Li, W.; Liang, Y.; Yu, D.; Peng, L.; Pernstich, K. P. Ultraviolet/ozone Treatment to Reduce Metal-graphene Contact Resistance. *Appl. Phys. Lett.* **2013**, *102*, 183110.

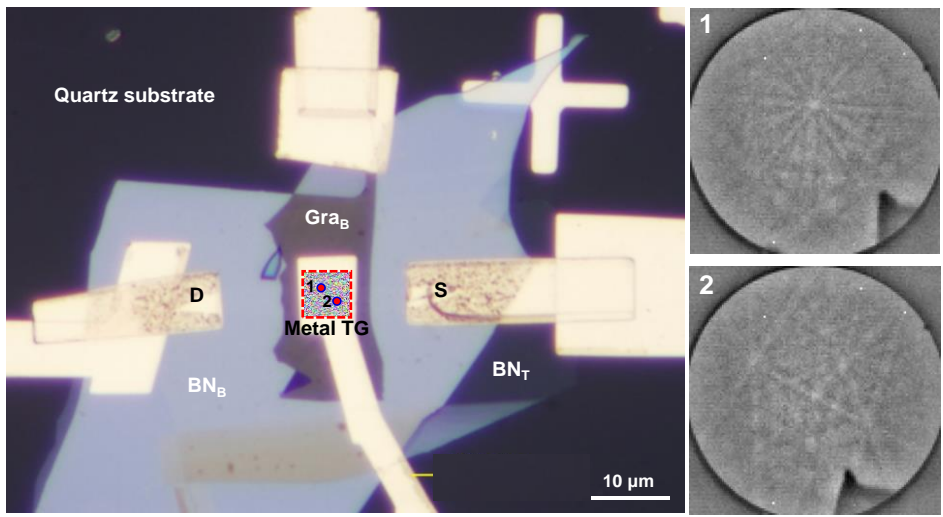

**Figure S4** Left: Optical image of $BN_T/BLG/BN_B/Gra_B$ device with metal top gate electrode on quartz substrate. The thicknesses of *h*-BN top gate, *h*-BN back gate, and graphite back gate electrode are 12 nm, 50 nm, 13.4 nm respectively. Right: EBSD diffraction pattern at different positions (1 and 2 in the left figure), indicating that the electrode metal is polycrystalline. This causes the potential fluctuation in BLG due to the variation of work function for different crystal orientation.

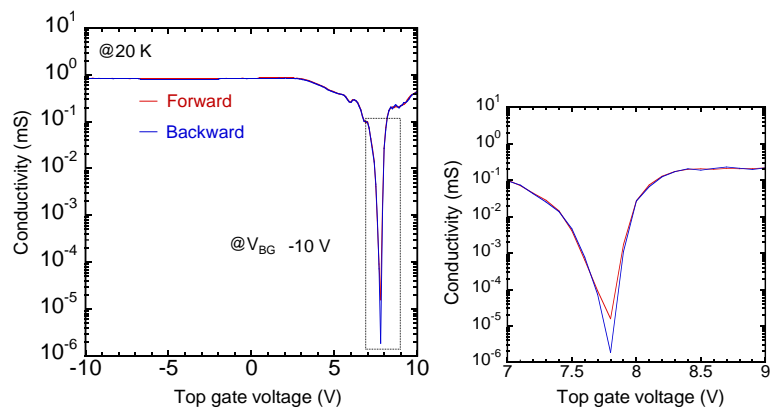

**Figure S5** Left: two-terminal conductivity as a function of forward and back ward top gate voltage sweeps for back gate voltage of -10 V measured at 20 K in Fig.2 (a). Right: enlarged conductivity in the doted area in the left figure.

**Note 2. Details of *C-V* measurement**

Lakeshore low temperature prober CRX-6.5K with closed-cycle cryostat system and Keysight E4980A LCR meter were used to measure the capacitance in this study. Generally, the parasitics of the prober system is removed by using the cable connection shown in **Fig. S6**. However, this cable connection increased the noise level because of the vibration from the mechanical refrigerator. Therefore, in this study, the parasitics were removed by performing open and short corrections before measuring the samples (the electrical correction), resulting in the noise floor of ~ 2 fF. However, the capacitance value that we measure is the order of several 10 fF due to the limited are of top gate in the heterostructure, even though *h*-BN is thin. In this case, the contribution from the metal electrode pad is quite large because of the large area. Therefore, the limiting factor is the residue frequency-dependent parasitics from the sample, not from the measurement system. These issues are critical for investigating the interface trap density by the conductance method, as frequency dependence resulted from the substrate is included in frequency dependence resulted from the interface traps. Therefore, the quartz substrate was used to eliminate these problems.

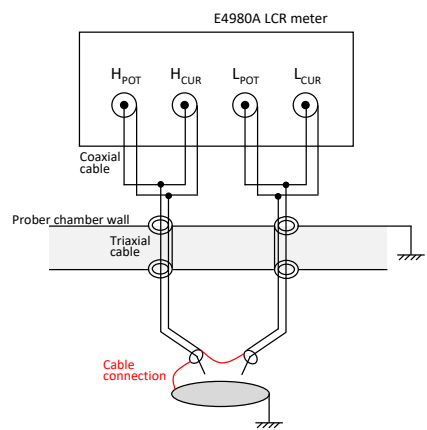

**Figure S6** Schematic representation of cable connection for removing parasitics in *C-V* measurement performed in this study.

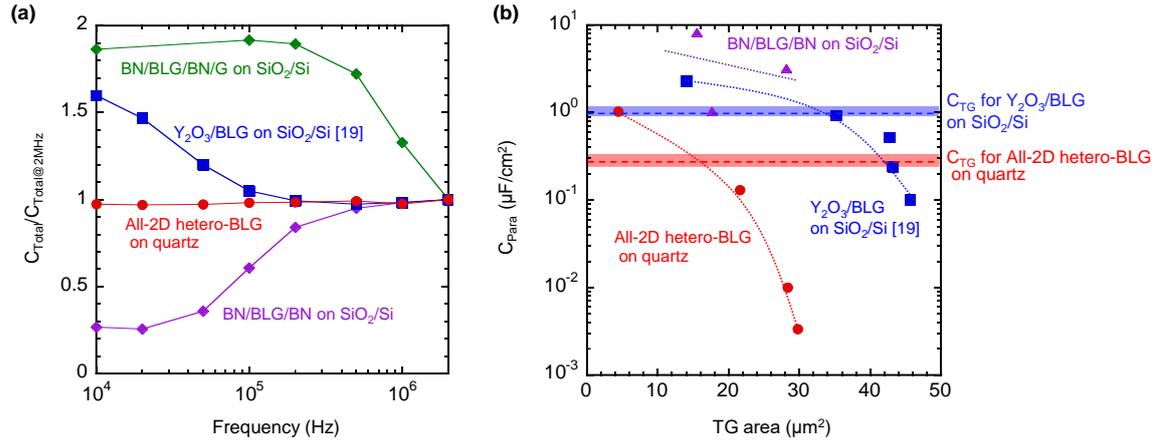

**Figure S7** (a) $C_{Total}$ as a function of frequency, normalized by that measured at 2 MHz. (b) Parasitic capacitance as a function of top grate area. The typical value for $C_{TG}$ is also shown by hatched region. In case of all-2D heterostructure BLG device on quartz, $C_{Para}$ is more than an order of magnitude smaller than $C_{TG}$, indicating $C_{Para}$ is negligible.

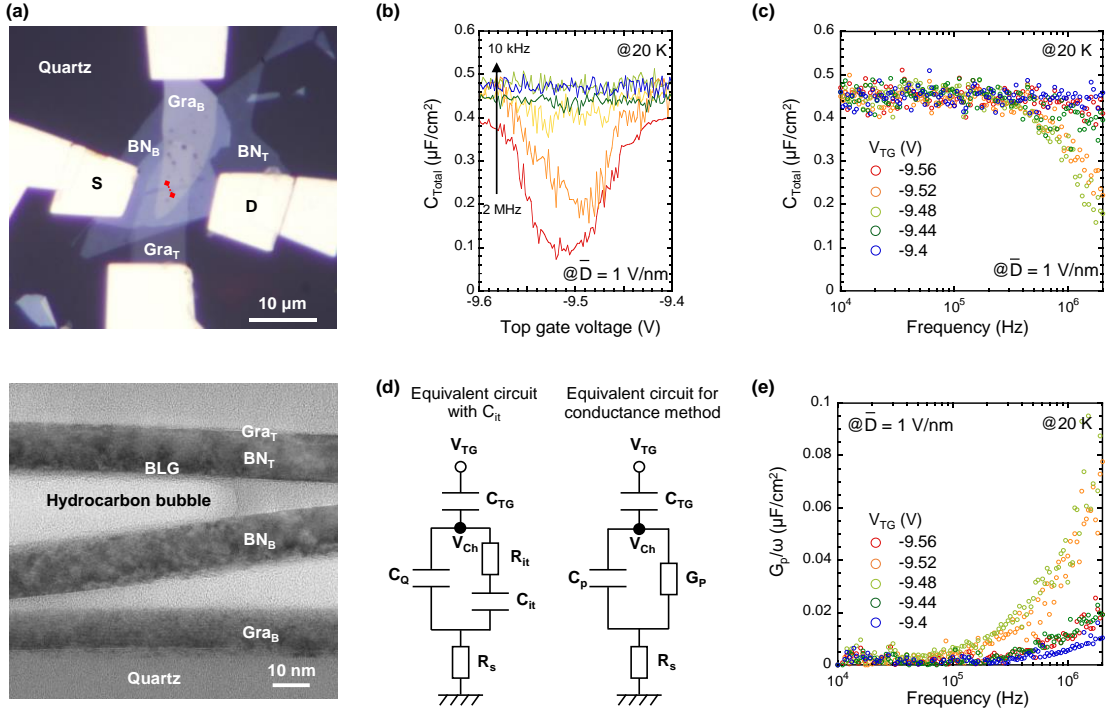

**Figure S8** (a) Optical image of all-2D-heterostructure BLG-FET with bubbles in the channel area. A cross section TEM image taken along the red dotted line in (a) is shown in the bottom. (b) $C_{Total}$ as a function of $V_{TG}$ measured at constant $\bar{D}$ = 1 V/nm in the device with bubbles in BLG channel at 20 K. It should be again noted that $V_{BG}$ is also changed to maintain constant $\bar{D}$. (c) $C_{Total}$ as a function of frequency extracted from the total impedance measured at each gate voltage in (b). (d) Left: Equivalent circuit including the contribution of interface traps in the device. $C_{it}$ and $R_{it}$ are capacitance and resistance associated with the interface traps, $R_S$ is series resistance. Right: The circuit can be converted into parallel capacitance and conductance ($C_p$ and $G_p$). $D_{it}$ is extracted from $D_{it}$ = $(G_p/\omega)_{max}/0.402q^2$, where $\omega = 2\pi f$ ($f$: measured frequency) and time constant ($\tau_{it}$) is obtained from $\tau_{it}$ = $1.98/2\pi f_0$, where $f_0$ is the frequency at $(G_p/\omega)_{max}$. (e) $G_p/\omega$ as a function of $f$ extracted from the total impedance measured at each gate voltage in (b). The peaks of $G_p/\omega$ seem to be in the frequency range higher than measurement limit in this study, suggesting that $\tau_{it}$ is shorter than measurement limit in this study, therefore $D_{it}$ could not be extracted quantitatively. However, the present result shows the response of $C_{it}$ which is consistent to the frequency response of $C_{total}$, supporting the interpretation of the data in **Fig. 3(c, d)**.